\documentclass[10pt]{article}
\usepackage{moriond,epsfig,epsf,colordvi}

\usepackage{amsmath,amsthm,amsfonts,amssymb,wrapfig}

\def\PRD{{\em Phys. Rev.} D}

\def\be{\begin{equation}}
\def\ee{\end{equation}}
\def\bea{\begin{eqnarray}}
\def\eea{\end{eqnarray}}

\begin{document}
\vspace*{4cm}
\title{CONSTRAINTS ON THE EXTRA DIMENSION BY KK GRAVITINO DECAY}

\author{ David Gherson }

\address{ Institut de Physique Nucl\'eaire de Lyon (IPNL), Universit\'e Lyon-I,\\
     Villeurbanne, France       }

\maketitle\abstracts{We study the consequences of the gravitino decay into dark matter. We suppose that the lightest neutralino is the main component of dark matter. In our framework gravitino is heavy enough to decay before Big Bang Nucleosynthesis starts. We consider a model coming from a five dimensional supergravity compactified on $S^1/Z_2$ with gravity in the bulk and matter localized on tensionless branes at the orbifold fixed points. We require that the dark matter, which is produced thermally and in the decay of Kaluza-Klein modes of gravitino, has an abundance compatible with observation. We deduce from our model that there are curves of constraints between the size of the extra-dimension and the reheating temperature of the universe after inflation. This talk is based on hep-ph/0702183 to be published in \PRD.}

\section{Introduction}

The five dimensional picture of the Universe has attracted much interest in the framework of what is called brane world cosmology. In the present work we choose to work in a set-up where we assume that the radion is stabilized in a five dimensional supergravity compactified on $S^1/Z_2$ where matter and gauge fields live on the branes and gravity in the bulk (\cite{Bagger:2001ep}, \cite{DeCurtis:2003hs}). Low energy supersymmetry provides a natural candidate for dark matter if R-Parity is conserved and solves the hierarchy problem.\\ 
As a general framework we can choose a scenario in which susy breaking is mediated to the observable sector partly by gravity anomaly and partly by Scherk-Schwarz mechanism. It is a scenario which can avoid the appearance of tachionic masses which are present in pure anomaly mediated scenario and problems due to pure gravity mediation scenario. This mechanism provides high masses for gravitino (it means that the mass of the gravitino can be above $10$ TeV).\\
The gravitino field has Kaluza-Klein excitations modes due to the presence of the extra-dimension. We suppose that all gravitino modes are produced after inflation during the reheating period by scattering effects in the primordial thermal bath. 
We suppose that the lightest mode - i.e. the zero mode - is heavy enough to mainly decay before the Big Bang Nucleosynthesis starts and we calculate its corresponding mass. Actually this is naturally the case in some scenarios of susy breaking (anomaly mediation or mix between anomaly and Scherk-Schwarz mechanism). Gravitinos modes decay into supersymmetric particles and standard model ones. If R-parity is conserved, all gravitino modes will give at the end of their decay cascade a Lightest Supersymmetric Particle (LSP) which is assumed to be the lightest neutralino. But not all of these decay products of gravitino will contribute to the relic density of neutralinos which is assumed to be the dark matter. Actually only the gravitino modes which decay after the thermal decoupling of neutralino contribute to the dark matter density. If a gravitino mode decays before the thermal decoupling of neutralinos it does not increase the number density of neutralinos since these gravitinos produce neutralinos which are in thermal equilibrium. So a finite number of gravitinos modes contribute to the non thermally produced dark matter. The total of the thermally produced and non thermally produced amount of dark matter is constrained by the evaluation of the dark matter content of the universe. As a consequence, we can draw curves of constraints between the size of the extra-dimension and the reheating temperature because the number of KK gravitinos modes is related to the size of the extra-dimension and the number density of gravitinos is related to the reheating temperature. We have chosen a reheating temperature in the range $10^5$ GeV to $10^{10}$ GeV. This range is quite natural for scenarios which allow baryogenesis through leptogenesis. There is another constraint on the size of the extra-dimension coming from the fact that there are also KK gravitons which can disturb BBN if the number of KK modes is too high. We checked with the help of the curves given by Jedamzik\cite{Jedamzik:2006xz} that for $R^{-1} >1$ TeV, this is not the case. 
  
\section{Interactions between gravitino and MSSM  }
We find (see(\cite{Gherson:2007uh})):
\begin{eqnarray}
{\cal L}_{interKK}^{4d}=\sum_{n=0}^{\infty} 
( -\frac{1}{\sqrt{2}M} e g_{ij^*} \tilde{\cal D}_{\nu} \phi^{*j}
\chi^i \sigma^\mu \bar{\sigma}^\nu \psi_{n,\mu}
-\frac{1}{\sqrt{2}M} e g_{ij^*} \tilde{\cal D}_{\nu} \phi^{i}
\overline{\chi}^j \bar{\sigma}^\mu \sigma^\nu \overline{\psi}_{n,\mu}
\nonumber          \\ 
-\frac{i}{2M} e \left(
\psi_{n,\mu} \sigma^{\nu\lambda} \sigma^\mu \overline{\lambda}_{(a)}
+ \overline{\psi}_{n,\mu} \bar{\sigma}^{\nu\lambda}
\bar{\sigma}^\mu \lambda_{(a)} \right)
F_{\nu\lambda}^{(a)})
\label{L4new}
\end{eqnarray}

Where $\phi$ are scalar fields, $\chi$ are chiral fermions, $\lambda$ are gauge fermions (gauginos) , $F_{\mu\nu}^{(a)}$ is the field strength tensor for the gauge boson $ A_\mu^{(a)}$. Indices i, j..represent species of chiral multiplets and (a), (b).... are indices for adjoint representation of gauge group. e is the vierbein. $ g_{ij^*}$ is the Kahler metric and $\psi_{n,\mu}$ is the KK gravitino field for the $n^{th}$ mode.

The masses of the modes n are related to the mass of the zero mode by the relation (\cite{Bagger:2001ep},\cite{DeCurtis:2003hs}):
\begin{equation}
M_{n}=M_0 + \frac{n}{R}
\label{mass}
\end{equation}

\section{Abundances and lifetime}
 Abundance for the zero mode is given by Kohri and al.\cite{Kohri:2005wn} taking into account production during the inflaton-dominated epoch and Pradler and al.\cite{Pradler:2006hh} have the same result without taking into account production during inflation. This result is given for masses of gravitino much higher than gauginos masses but the calculation of the creation term in the Boltzmann equation is made with particles without masses: their mass is supposed negligible compared to the average energy in the center of mass of each reaction. The formula is :
\begin{eqnarray}
      Y_{3/2} &\simeq& 
    1.9 \times 10^{-12}
    \nonumber \\ &&
    \times \left( \frac{T_{\rm R}}{10^{10}\ {\rm GeV}} \right)
    \left[ 1 
        + 0.045 \ln \left( \frac{T_{\rm R}}{10^{10}\ {\rm GeV}} 
        \right) \right]
    \left[ 1 
        - 0.028 \ln \left( \frac{T_{\rm R}}{10^{10}\ {\rm GeV} } 
        \right) \right],
    \nonumber \\
\label{Ygrav}
\end{eqnarray}

Where $Y_{3/2}=\frac{n_{3/2}}{s}$, $n_{3/2}$ is the number density, $s$ is the entropy density and $T_R$ the reheating temperature.
The quantity $Y=\frac{n}{s}$ is the density per comoving volume. We have to take into account in our calculation the different masses of gravitino modes. We find as a good approximation this rule for the abundance of the different modes:
\begin{eqnarray}
Y_{3/2}^k = Y_{3/2}^0 \ ,& {\rm for}\  M^k\le T_{\rm R}& \rm{and}\nonumber\\
Y_{3/2}^k = 0\ ,& {\rm for}\ M^k > T_{\rm R}
\label{abondancekk}
\end{eqnarray}
Where k represents the index of the KK mode, $M^k$ is the mass of the $k^{th}$ gravitino mode and $Y_{3/2}^k$ its abundance. \\
We have calculated the life time for heavy gravitino (${\rm masses} > 10\ {\rm TeV}$) using (\cite{Kohri:2005wn}):
\begin{eqnarray}
\tau_k = 1.4\ 10^7\times \left( \frac{M_k}{100 {\rm GeV}} \right)^{-3} \ {\rm Sec}
\label{tempsdevie}
\end{eqnarray}

\section{Neutralinos and equation of constraints}
In our model the LSP is the lightest neutralino. We choose to work with a mass of LSP equal to $120$ GeV. This analysis can be easily rescaled for another choice for the mass: we also show results for a mass of the LSP equal to $200$ GeV in the article\cite{Gherson:2007uh} . The dark matter density is:
\begin{eqnarray}
0.106\ <\Omega\ h^2\ <\ 0.123 
\label{matiere noire}
\end{eqnarray}

We call  $\Omega_{th}$ the thermal density of neutralinos. We find this approximate relation between $\Omega_{th}$ and $x_f$ :
\begin{eqnarray}
\Omega_{th}\ h^2 = 3.61\ 10^6 \ \frac{m_{lsp}}{1 \rm{GeV}}\ x_f^2\ e^{-x_f}
\label{omegath}
\end{eqnarray}

We choose different values of $\Omega_{th} \ h^2$ and complete with the non thermal production coming from the gravitino decay. We call this non thermal production $\Delta \Omega \ h^2$. 
\begin{eqnarray}
0.106 \le \Omega_{th}\ h^2 + \Delta\Omega \ h^2 \le 0.123
\label{encadrement}
\end{eqnarray}

 As one gravitino produces one neutralino, we can write:
\begin{equation}
\Delta\Omega \ h^2 =\frac{ m_{lsp}\ s_0\ h^2}{\rho_c} \ \sum\limits_{k=0}^{k=\bar{n}} Y_{3/2}^k
\label{deltaomega}
\end{equation}
The index $\bar{n}$ corresponds to the last mode to be taken into account. It is the mode decaying just when the LSP decouples from the thermal bath. Only the gravitino modes decaying after the thermal decoupling of the neutralinos contribute to increase the quantity of neutralinos. 

Using the equations (\ref{Ygrav}) and (\ref{abondancekk}), and the equation (\ref{deltaomega}) we find a constraint equation coming from the relation (\ref{encadrement}):
\begin{eqnarray}
 & &R^{-1}=\left( M^{\bar{n}}-M^0 \right)\times
\nonumber\\
 & &\left( I\left[\Delta\Omega \ h^2 \frac{\rho_c}{m_{lsp} \ s_0 \ h^2}\ \frac{1}{1.9 \times 10^{-12}\times  \frac{T_{\rm R}}{10^{10}} \left[ 1 + 0.045 \ln \left( \frac{T_{\rm R}}{10^{10}}\right) \right]\left[ 1 - 0.028 \ln \left( \frac{T_{\rm R}}{10^{10}}\right) \right]}\right]-1 \right)^{-1}
\nonumber\\
\label{final}
\end{eqnarray}

\section{Results}
In the paper(\cite{Gherson:2007uh}), we have treated three cases for different values of $\Omega_{th}\ h^2$. As an example we present below two figures of the case where $\Omega_{th}\ h^2= 0.053 $, $(\Delta\Omega_{th} \ h^2)_{min}= 0.053$, $(\Delta\Omega_{th} \ h^2)_{max}= 0.070 $, $M_{\bar{n}}= 9.66 \times 10^6$ GeV.

\begin{figure}[h]
\centerline{\includegraphics[height=7cm]{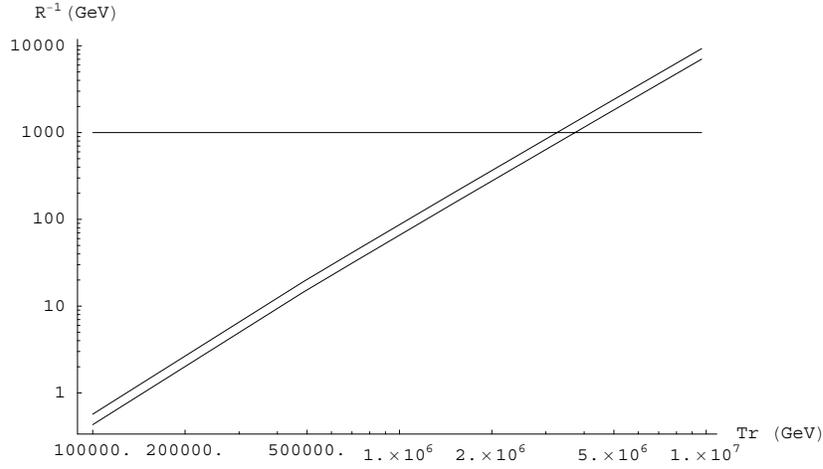}} 
\caption{\label{fig1}$T_R$ less than $9.66 \ 10^6$ GeV. Only the band between the two diagonal curves is allowed. The zone below the straight line $R^{-1}=1$ TeV is excluded if KK gravitons constraint is taken into account.}
\end{figure}

\begin{figure}[h]
\centerline{\includegraphics[height=7cm]{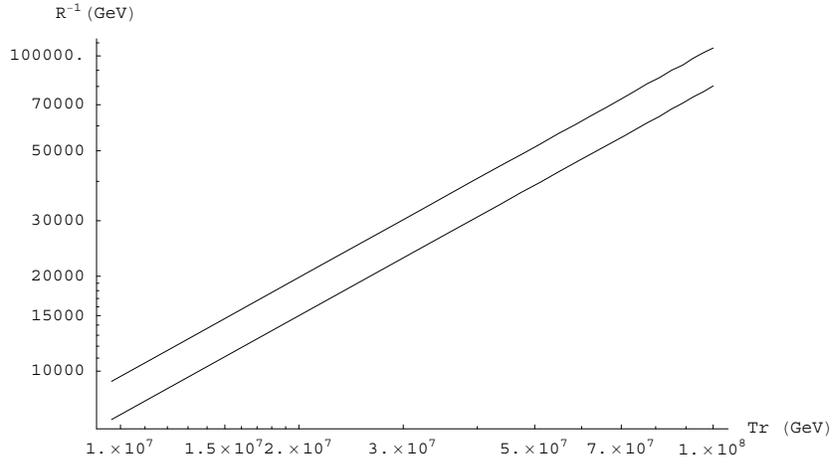}}
\caption{\label{fig2} $9.66 \ 10^6 \ GeV \le T_R \le  10^8 \ GeV $. Only the band between the two curves is allowed.}
\end{figure}

\section{Conclusion}
 The results that we obtain are independent from the susy mass spectrum since the gravitino is heavy enough to make negligible the influence of other susy particles. The results show that the size of the radius R is not only bounded by a maximum value but also by a minimum value in a wide range of possible values for the thermal production of neutralinos and in a wide range of values for the reheating temperature. To obtain a minimum size for the radius is a new result. \\
Kaluza-Klein modes of graviton are also present and may disturb BBN. We checked with an approximate method that for $R^{-1}$ above 1 TeV it is not the case. This already implies a bound on the reheating temperature which can not be lower than a minimum value in the cases where the radius R is bounded by a minimum value.

\section*{Acknowledgments}

I thank very much Aldo Deandrea for his contribution to this work, Sacha Davidson, Karsten Jedamzik and the referee of my paper\cite{Gherson:2007uh} for useful discussions and comments.

\end{document}